\documentclass[12pt,preprint,linenumbers]{aastex}
\usepackage{lineno}

\shorttitle{magnetic parameters of large solar flares} \shortauthors{Li et
al.}

\begin{document}

\title{Survey of Magnetic Field Parameters Associated With Large Solar Flares}

\author{Ting Li\altaffilmark{1,2,3}, Yanfang Zheng\altaffilmark{4}, Xuefeng Li\altaffilmark{4}, Yijun Hou\altaffilmark{1,2,3}, Xuebao Li\altaffilmark{4}, Yining Zhang\altaffilmark{1,2} \& Anqin Chen\altaffilmark{5}}

\altaffiltext{1}{CAS Key Laboratory of Solar Activity, National
Astronomical Observatories, Chinese Academy of Sciences, Beijing
100101, China; liting@nao.cas.cn} \altaffiltext{2}{School of
Astronomy and Space Science, University of Chinese Academy of
Sciences, Beijing 100049, China} \altaffiltext{3}{National Space Science Center, Chinese Academy of Sciences, Beijing 100190, China}
\altaffiltext{4}{School of computer science, Jiangsu University of Science and Technology, Zhenjiang, China; zyf062856@163.com}
\altaffiltext{5}{Key Laboratory of
Space Weather, National Center for Space Weather, China
Meteorological Administration, Beijing 100081, China}

\begin{abstract}

Until now, how the magnetic fields in M/X-class flaring active regions (ARs) differ from C-class flaring ARs remains unclear. Here, we calculate the key magnetic field parameters within the area of high photospheric free energy density (HED region) for 323 ARs (217 C- and 106 M$/$X-flaring ARs), including total photospheric free magnetic
energy density E$_{free}$, total unsigned magnetic flux $\Phi$$_{HED}$, mean unsigned current helicity h$_{c}$, length of the polarity inversion lines $L$$_{PIL}$ with a steep horizontal magnetic gradient, etc., and compare these with flare/coronal mass ejection (CME) properties. We first show the quantitative relations
among the flare intensity, the eruptive character and $\Phi$$_{HED}$. We reveal that $\Phi$$_{HED}$ is a measure for the GOES flux upper limit of the flares in a given region. For a given $\Phi$$_{HED}$, there exists the lower limit of F$_\mathrm{SXR}$ for eruptive flares. This means that only the relatively strong flares with the large fraction of energy release compared to the total free energy are likely to generate a CME. We also find that the combinations of E$_{free}$-$L$$_{PIL}$ and E$_{free}$-h$_{c}$ present a good ability to distinguish between C-class and M$/$X-class flaring ARs. Using determined critical values of E$_{free}$ and $L$$_{PIL}$, one predicts correctly 93 out of 106 M/X-class flaring ARs and 159/217 C-class flaring ARs. The large $L$$_{PIL}$ or h$_{c}$ for M$/$X-class flaring ARs probably implies the presence of a compact current with twisted magnetic fields winding about it.

\end{abstract}


\section{Introduction}

Solar flares and coronal mass ejections (CMEs) are powerful explosive events in the solar atmosphere, which can
cause severe space whether disturbances. During these events, sudden huge magnetic energy stored in active regions (ARs) is converted to plasma
energy through magnetic reconnection (Priest \& Forbes 2002; Fletcher et al. 2011; Li et al. 2021a). The non-potentiality of ARs can be characterized by a variety of parameters, such as magnetic free energy (Su et al. 2014; Kusano et al. 2020), magnetic shear (Hagyard et al. 1984; Chen et al. 2021), electric currents (Georgoulis et al. 2012; Avallone \& Sun 2020), magnetic helicity (Tziotziou et al. 2012; Zuccarello et al. 2018; Liu et al. 2023) and so on. Historically, most of the
analyses of photospheric magnetic field parameters focused on the AR as a whole (Leka \& Barnes 2003; Toriumi
\& Wang 2019). But actually, the magnetic field structure near core regions of ARs plays an important role in determining the flare occurrence (Schrijver 2007; Sun et al. 2015). Moreover, the quantitative studies have shown that only
a fraction of the AR magnetic field is involved in the magnetic reconnection of the flare (Kazachenko et al. 2017; Tschernitz et al. 2018; Li et al. 2020). These imply that the attempt to reveal the discriminants between flare/CME-producing and flare/CME-quiet regions requires our understanding of the magnetic properties of AR core regions.

Solar flares are often, but not always,
accompanied by CMEs. The flares associated with a CME are named as ``eruptive flares"
and flares not associated with a CME as ``confined flares" within this work. Previous studies have shown that there are two factors determining the eruptive character of solar flares. The first factor describes the constraining effect of the overlying
field, e.g., its decay rate with height and its strength (T{\"o}r{\"o}k \& Kliem 2005; Wang \& Zhang 2007; Cheng et al. 2011; Amari et al. 2018; Baumgartner et al. 2018; Duan et al. 2019; Li et al. 2020, 2021b). The second factor is the degree of the AR non-potentiality (Nindos \& Andrews 2004; Cui et al. 2018; Liu et al. 2018; Vasantharaju et al. 2018; Thalmann et al. 2019; Wang et al. 2023), i.e., magnetic helicity, twist, shear, etc. Recently, several studies have shown that the relative structural relation between the magnetic fields of the flaring region and the surrounding
magnetic structures may determine the eruptive character of solar flares (Pariat et al. 2017; Gupta et al. 2021; Lin et al. 2021). Li et al. (2020) and Kazachenko et al. (2022) revealed that confined flares have smaller fractions of the AR magnetic flux and area that participate in the flare. Li et al. (2022) found that the ratio of the magnetic twist within
the flaring polarity-inversion line (PIL) to the AR magnetic flux can well distinguish
confined from eruptive events.

Based on the classic Space-Weather HMI AR Patches (SHARP; Bobra et al. 2014) descriptors calculated from vector and line-of-sight magnetograms, there have been numerous flare forecasting studies specifically targeted the forecasting of the M/X-class flares (Bobra \& Couvidat 2015; Barnes et al. 2016; Liu et al. 2017; Huang et al. 2018; Campi et al. 2019; Zheng et al. 2023). However, these previous studies are still not capable of providing a substantially better performance than climatological forecasts (Barnes et al. 2016; Campi et al. 2019), which implies that the most relevant physical parameters that facilitate the prediction of solar flares are still unknown (Aschwanden 2020; Kontogiannis 2023). In this study, we introduce the area of high photospheric magnetic free energy
density (HED) as the proxy of AR core region and analyze 11 non-potential parameters within HED areas of 323 ARs producing C-class and M/X-class flares. We demonstrate that individually these parameters do not have a
good ability to separate between C-flaring ARs and M/X-flaring ARs, but in certain combinations, these two groups of ARs can be distinguished. Here, C-flaring AR means an AR that produced only C-class flares (no M/X-class flares) while being within $45^{\circ}$ from the solar disk center, and M/X-flaring AR denotes an AR that produced at least one M/X-class flare while being within $45^{\circ}$ from the solar disk center.

\section{Database Selection and Parameter Calculations}

Here, we establish a catalog of 323 ARs appearing from June
2010 until December 2022 observed by the Helioseismic and Magnetic Imager
(HMI; Scherrer et al. 2012) on the Solar Dynamics Observatory (\emph{SDO}; Pesnell et al. 2012). The criteria we use to assemble
our catalog are that (1) the
ARs produced at least one C-class flare within $45^{\circ}$ from the central meridian. (2)
During 30 min before the largest flare onset that the ARs produced within $45^{\circ}$, there exist the areas of HED (photospheric free energy
density $\rho_\mathrm{free}$$\geq$2.0$\times$10$^{4}$ erg cm$^{-3}$) in the AR. About 63\% C-flaring ARs (378 in 595) and 23\% M/X-flaring ARs (32 in 138) are excluded for the absence of HED regions. The second criterion was used because we calculate the magnetic parameters within the HED region which is thought to be closely related with flaring regions. Finally, a total of 217 C- and 106 M$/$X-flaring ARs are selected. For each selected AR, we focus on its maximum GOES magnitude of flares they generated within $45^{\circ}$.
To determine whether a flare is associated
with a CME, we use the database
FlareC5.0\footnote{\url{http://dx.doi.org/doi:10.12149/101067}} from Li et al. (2021b) and the CME catalog\footnote{\url{https://cdaw.gsfc.nasa.gov/CME\_list/}} of the Solar and
Heliospheric Observatory/Large Angle and Spectrometric
Coronagraph (Brueckner et al. 1995).

We use the vector magnetograms from SHARP before the flare onset and calculate 11 photospheric magnetic field parameters. For each AR, the length of the
PILs ($L$$_{PIL}$) with a steep horizontal magnetic gradient ($\geq$ 300 G Mm$^{-1}$) is calculated over the entire AR. Different from $L$$_{PIL}$, 10 other magnetic parameters (total photospheric free magnetic
energy density E$_{free}$, total unsigned magnetic flux $\Phi$$_{HED}$, area of HED $S$$_{HED}$, mean unsigned vertical current density J$_{z}$, total unsigned vertical current J$_{total}$, mean unsigned current helicity h$_{c}$, total unsigned
current helicity h$_{total}$, mean shear angle $\Psi$, mean characteristic twist parameter $\alpha$ and mean horizontal gradient of horizontal field $\nabla$B$_{h}$) are calculated only within the HED region. Detailed formulas of the parameters are listed in Table 1. Here we create a new data
set, MagParDB\footnote{\url{http://dx.doi.org/doi:10.12149/101362}}, and describe the above 11 parameters within HED areas.

\section{Statistical Results}

Fig. 1 shows an example of AR 12205 producing an X1.6-class flare
that started at 16:53 UT and peaked at 17:26 UT on 2014 November 7. It can be seen that this AR accumulated a large amount of photospheric free magnetic energy density E$_{free}$ reaching about 4.0$\times$$10^{23}$ erg cm$^{-1}$ within the HED region (orange contours in Figures 1(a)-(d)). Figure 1(a) shows that the HED region just surrounds the PILs with a steep horizontal magnetic gradient. Within the HED region, there is strong and concentrated current helicity along both sides of the PILs, indicating the presence of a
highly sheared or twisted field lines (Figure 1(c)). The HED region approximately corresponds to the initial ribbon brightenings at 1600 {\AA} (Figure 1(d)), which implies that the HED region can represent the flare-related AR core area.

We make a survey of the relations between calculated 11 magnetic parameters for 323 ARs and the corresponding flare/CME properties.
These flaring ARs are classified into four types: eruptive M/X-class, confined M/X-class, eruptive C-class and confined C-class. The scatter diagram relating the GOES 1-8 {\AA} peak flux, F$_\mathrm{SXR}$, with total unsigned magnetic flux $\Phi$$_{HED}$ within HED region for the four types of ARs is shown in Figure 2(a). We can see that $\Phi$$_{HED}$ is a measure for the GOES flux upper limit in a given region (F$_\mathrm{SXR}$$\leq$1.26$\times$$10^{-14}$$\Phi$$_{HED}$$^{0.5}$; black solid line in Figure 2(a)). That is, ARs can reduce their free energy by relaxing through a series of flares and the maximum GOES magnitude of flares can be estimated based on the value of $\Phi$$_{HED}$. For example, for an AR with $\Phi$$_{HED}$$\geq$5.3$\times$$10^{19}$ Mx, an X-class flare is likely to occur (black dotted lines in Figure 2(a)). Moreover, there exists a lower limit of eruptive flares (F$_\mathrm{SXR}$$\leq$4.79$\times$$10^{-17}$$\Phi$$_{HED}$$^{0.54}$; orange line in Figure 2(a)) above which the flares are likely to be associated with CMEs. If the flare peak intensity is below the limit, this flare is a confined flare regardless of C-class or M/X-class flares. The larger $\Phi$$_{HED}$ an AR has, the stronger flare is needed to generate a CME. For instance, for an AR with $\Phi$$_{HED}$=1.0$\times$$10^{21}$ Mx, the flares larger than M2.5-class are possible to produce a CME. It needs to be noted that for each AR only the largest flare that occurred in the AR within $45^{\circ}$ from the solar disk center is considered. In the subregion below the limit (the orange line), other smaller flares in an AR are probably confined because the flare-CME association is steeply decreasing with the decreasing flare intensity (Yashiro et al. 2006; Li et al. 2021b).

The diagram relating F$_\mathrm{SXR}$ with total photospheric free magnetic energy density E$_{free}$ within HED region shows similar results (Figure 2(b)).
The black solid line at F$_\mathrm{SXR}$$=$1.91$\times$\\
$10^{-19}$E$_{free}$$^{0.66}$ has 99\% of the observed flares below it (three flares slightly lie above this line).
To generate an X1.0-class flare, E$_{free}$ at least reaches the value of 1.7$\times$$10^{22}$ erg cm$^{-1}$. At the bottom right corner, all the observed flares except one are confined, which are below the orange lie at F$_\mathrm{SXR}$$=$1.38$\times$$10^{-20}$E$_{free}$$^{0.64}$.

In order to reveal the magnetic-field properties that distinguish M/X-flaring regions and C-flaring regions, we make certain combinations of magnetic parameters. Here, we show the outcomes of the classification problem in the confusion matrix: we classified as true positives (TPs) all M/X-flaring ARs that have been correctly predicted as M/X-class flaring; as true negatives (TNs) all C-flaring ARs that have been correctly predicted as C-class flaring; as false negatives (FNs) all M/X-flaring ARs that have been incorrectly predicted as C-class flaring; and as false positives (FPs) all C-flaring ARs incorrectly predicted as M/X-flaring. The E$_{free}$$-$PIL length diagram in Figure 3(a) (hereafter E$_{free}$-$L$$_{PIL}$ diagram) shows that a large majority of M/X-flaring ARs lie in region (i) (E$_{free}$$\geq$2.5$\times$$10^{22}$ erg cm$^{-1}$ and L$_{PIL}$$\geq$12 Mm) and most of C-flaring ARs lie in regions (ii)-(iv). In the confusion matrix, TPs equal 93, TNs equal 159, FNs equal 13 and FPs equal 58. In the subregion with E$_{free}$$\geq$2.0$\times$$10^{23}$ erg cm$^{-1}$ and L$_{PIL}$$\geq$35 Mm (outlined by green dash-dotted lines), the largest possible flare that the AR is capable of producing is $\geq$M1.0-class. The vast majority of ARs lie in regions (i) and (iii), which contain populations with high (low) magnetic energy and long (short) high-gradient PILs, respectively. This reflects that in a statistical sense, most ARs that are highly charged with magnetic energy have long high-gradient PILs. E$_{free}$ and $L$$_{PIL}$ are highly correlated
with each other with the Spearman rank order
correlation coefficient r$_{s}$ of 0.82. The least-squares best logarithmic fit is
\begin{equation}
\log{L_{PIL}}=(0.42\pm0.02)\log{E_{free}}+(-8.17\pm0.51). \label{eq1}
\end{equation}

The $\Phi$$_{HED}$-$L$$_{PIL}$ diagram shows an overall trend under which both $\Phi$$_{HED}$ and $L$$_{PIL}$ increase together (Figure 3(b)). Similar to the E$_{free}$-$L$$_{PIL}$ diagram, large flares tend to occur in ARs with $\Phi$$_{HED}$$\geq$$10^{20}$ Mx and L$_{PIL}$$\geq$12 Mm. The least-squares best logarithmic fit between $\Phi$$_{HED}$ and L$_{PIL}$ reveals a scaling of the form
\begin{equation}
\log{L_{PIL}}=(0.42\pm0.02)\log{\Phi_{HED}}+(-7.29\pm0.43). \label{eq1}
\end{equation}

The E$_{free}$$-$mean current helicity diagram in Figure 3(c) (hereafter E$_{free}$-h$_{c}$ diagram) shows that TPs equal 91, TNs equal 153, FNs equal 15 and FPs equal 64 in the confusion matrix by using
the thresholds of E$_{free}$$=$2.5$\times$$10^{22}$ erg cm$^{-1}$ and h$_{c}$$=$0.006 G$^{2}$ m$^{-1}$. If an AR exhibits characteristic values of E$_{free}$$\geq$2.0$\times$$10^{23}$ erg cm$^{-1}$ and h$_{c}$$\geq$0.015 G$^{2}$ m$^{-1}$, then the AR is most likely to produce the largest flare $\geq$M1.0-class (outlined by green dash-dotted lines). The E$_{free}$$-$mean shear angle diagram (E$_{free}$-$\Psi$ diagram) shows a similar pattern with E$_{free}$-h$_{c}$ diagram with the threshold of $\Psi$$\sim$ 45$^{\circ}$ (Figure 3(d)). In order to confirm the statistical results calculated in 30 min before the flare onset, we further analyze these parameters calculated about 24 hr before the flare, which show consistent results with Figure 3.

Figure 4 shows the scatter plots of $\Phi$$_{HED}$ vs. E$_{free}$, area of HED region S$_{HED}$ vs. E$_{free}$, total vertical current J$_{total}$ vs. E$_{free}$ and total current helicity h$_{total}$ vs. E$_{free}$. We can see that $\Phi$$_{HED}$ shows a strong correlation with E$_{free}$
with the Spearman rank order correlation coefficient r$_{s}$ of 0.91. Their relation is

\begin{equation}
\log{\Phi_{HED}}=(0.90\pm0.03)\log{E_{free}}+(-0.085\pm0.668). \label{eq1}
\end{equation}

It is natural that S$_{HED}$ and E$_{free}$ are strongly correlated with r$_{s}$ of about 0.94, and their least-squares best logarithmic fit is

\begin{equation}
\log{S_{HED}}=(0.72\pm0.02)\log{E_{free}}+(1.66\pm0.48). \label{eq1}
\end{equation}

The maximum correlation coefficient r$_{s}$ here is $\sim$0.99 between J$_{total}$ and E$_{free}$,

\begin{equation}
\log{J_{total}}=(0.90\pm0.007)\log{E_{free}}+(-8.47\pm0.16). \label{eq1}
\end{equation}

The total unsigned current helicity h$_{total}$ also shows a strong correlation with E$_{free}$ at r$_{s}$ of about 0.94, and their relation is

\begin{equation}
\log{h_{total}}=(1.07\pm0.02)\log{E_{free}}+(-23.05\pm0.48). \label{eq1}
\end{equation}

\section{Summary and Discussion}

In this paper, in order to improve our understanding of the physical properties of photospheric vector magnetic fields in C-flaring and M/X-flaring active regions, we analyzed the preflare vector magnetic fields
within high-$\rho_{free}$ areas in 323 ARs (217 C-class and 106 M$/$X-class
flaring ARs).
We find that for a given region, the maximum flare intensity is proportional to $\Phi$$_{HED}$ and can be determined by measuring $\Phi$$_{HED}$. For regions only with $\Phi$$_{HED}$$\geq$5.3$\times$$10^{19}$ Mx, an X-class flare is likely to occur. The lower limit of flare intensity to generate a CME is first shown in our study (4.79$\times$$10^{-17}$$\Phi$$_{HED}$$^{0.54}$), above which the flares are likely to be associated with CMEs. If the flare peak intensity is below the limit, this flare is a confined flare. The larger $\Phi$$_{HED}$ an AR has, the stronger flare is needed to generate a CME. The diagram relating F$_\mathrm{SXR}$ with E$_{free}$ shows similar results with the $\Phi$$_{HED}$-F$_\mathrm{SXR}$ diagram. Our statistical results also reveal that the combination of E$_{free}$ and L$_{PIL}$ presents a good ability to distinguish between C-class and M$/$X-class flaring ARs. A large majority of M/X-flaring ARs have E$_{free}$$\geq$2.5$\times$$10^{22}$ erg cm$^{-1}$ and L$_{PIL}$$\geq$12 Mm, and most of C-flaring ARs have smaller E$_{free}$ or shorter PILs. In the confusion matrix, TPs equal 93, TNs equal 159, FNs equal 13 and FPs equal 58.
The E$_{free}$-h$_{c}$ diagram shows similar results with E$_{free}$-L$_{PIL}$ diagram, with TPs of 91, TNs of 153, FNs of 15 and FPs of 64 in the confusion matrix by using
the thresholds of E$_{free}$$=$2.5$\times$$10^{22}$ erg cm$^{-1}$ and h$_{c}$$=$0.006 G$^{2}$ m$^{-1}$.

The source of free magnetic energy (energy above the current-free magnetic fields) is electric currents in the corona. The HED region at the photosphere corresponds to the area with a high electric current density and a large current helicity (Figure 1), which is closely related with the later flaring region. We find that $\Phi$$_{HED}$ is a measure for the GOES flux upper limit of the flares in a given region. Schrijver (2007) proposed the total unsigned flux $\emph{R}$ within about 15 Mm of strong-field, high-gradient PILs and revealed that the maximum flare is proportional to the value of R. The two parameters $\Phi$$_{HED}$ and $\emph{R}$ show similar results, probably because that the HED region and the area used to calculated R are in fact proxies of photospheric electrical currents. Aschwanden (2020) predicted the upper limit of the possible GOES class based on the observed scaling of the slowly varying potential energy in the region. The free energy and potential energy show a good positive correlation (see Figure 1 in Aschwanden 2020). In our study, E$_{free}$ can be used to estimate the strongest flare in an AR, which is consistent with the results of Aschwanden (2020).

Moreover, we first show the quantitative relations among the flare intensity, the eruptive character and $\Phi$$_{HED}$. We find that for a given $\Phi$$_{HED}$, there exists the lower limit of F$_\mathrm{SXR}$ for eruptive flares. This means that only the relatively strong flares with the large fraction of energy release compared to the total free energy are likely to generate a CME. Earlier, some indications of this
relationship were suggested by Li et al. (2020) and Kazachenko et al. (2022), who found that eruptive flares have, on average, larger fractions of AR magnetic flux participating in the flare. Lin et al. (2021) proposed the ratio of the magnetic flux of twist higher than a threshold to the overlying magnetic flux can provide a moderate ability for distinguishing the eruptive and confined events. The statistical results of Li et al. (2021b) show that the slope of the flare-CME association rate reveals a steep monotonic decrease with increasing $\Phi$$_{AR}$, implying that a large magnetic flux tends to confine eruptions. We suggest that $\Phi$$_{HED}$ is probably positively correlated with $\Phi$$_{AR}$, which describes the strength of the background field confinement (Li et al. 2020, 2021b). In this work, with the increasing $\Phi$$_{HED}$, the stronger flares are needed to generate a CME. Stronger flares usually correspond to larger upward force that drives the eruptions. Our results imply that the balance between the upward force and the downward force that suppresses the eruptions largely determines the eruptive character of flares.

We find that large flares are associated with both a large E$_{free}$ and a long $L$$_{PIL}$. Previous studies have shown that the free magnetic energy stored in an AR is a relatively good parameter closely related with large flares (Emslie et al. 2012; Liokati et al. 2022; Xu et al. 2022).
It has also been found that large flares often occur near strong and highly sheared PILs (Falconer et al. 2002; Sadykov \& Kosovichev 2017; Vasantharaju et al. 2018; Dhakal \& Zhang 2024). However, the importance of the combination of two parameters has never been demonstrated in previous studies. The combination of E$_{free}$ and $L$$_{PIL}$ shows a better ability to differentiate between C-class and M$/$X-class flaring ARs than only E$_{free}$ or $L$$_{PIL}$ considered. The E$_{free}$-h$_{c}$ diagram shows similar results with E$_{free}$-L$_{PIL}$ diagram. Our results imply that large solar flares occur because of the presence of a compact current with magnetic fields winding helically about it.

It needs to be noted that for each AR only the largest flare that occurred in the AR within $45^{\circ}$ from the solar disk center is considered, not for the entire population of flares (a total of 4743 flares $\geq$ C1.0-class from 733 ARs are recorded from June 2010 until December 2022 within 45 degrees from the solar disk center. Among these 733 ARs, about 63\% C-flaring ARs and 23\% M/X-flaring ARs are excluded for the absence of HED regions). Thus our results in this study are for the largest possible flare that the AR is capable of producing. Moreover, about 30\% of C-class flaring ARs have E$_{free}$ and $L$$_{PIL}$ beyond critical values (2.5$\times$$10^{22}$ erg cm$^{-1}$ and 12 Mm). This implies that in some ARs, only a small fraction of free energy is released and thus only C-class flares are generated. This raises a question that which factor determines the fraction of energy release for an AR. For ARs with similar E$_{free}$ or $L$$_{PIL}$, why some ARs produce large flares while others produce only C-class flares? In future, we need to extrapolate the three-dimensional coronal magnetic fields and further investigate the difference of magnetic fields in C-class flaring ARs and M/X-class flaring ARs.

\acknowledgments {This work is supported by the B-type Strategic Priority
Program of the Chinese Academy of Sciences (Grant No. XDB0560000 and XDB41000000), the
National Natural Science Foundations of China (12222306, 12273060), Natural Science Foundation of Jiangsu Province of China (Grants  No. BK20201199), the Qing Lan Project of Jiangsu Province, the Youth Innovation
Promotion Association of CAS (2023063) and the National Key R\&D
Program of China (2019YFA0405000). A. Q. Chen is supported by the
Strategic Priority Program on Space Science, Chinese Academy of
Sciences, Grant No. XDA15350203. \emph{SDO}
is a mission of NASA's Living With a Star Program.}

{}
\clearpage

\begin{table*}
\centering \caption{10 Parameters Calculated Within HED Region \label{tab1}} \centering
\begin{tabular}{c c c c c c c c c} 
\hline\hline 
Parameters & Description & Unit &
Formula \\ 
\hline 
$E_{free}$ & Total photospheric free & erg cm$^{-1}$ & E$_{free}$=$\Sigma$$\rho$$_{free}$dA\tablenotemark{\emph{1}}\\
  & energy density &   &  \\
$\Phi$$_{HED}$ & Total unsigned magnetic flux & Mx & $\Phi$$_{HED}$=$\Sigma$$|$$B_{z}$$|$dA\\
  & within HED region &   &  \\
S$_{HED}$ & Area within HED & cm$^{2}$ & S$_{HED}$=$\Sigma$dA\\
  & region &   &  \\
J$_{z}$ & Mean unsigned vertical & mA m$^{-2}$ & J$_{z}$=$\frac{1}{N\mu}$$\Sigma$$|$$\frac{\partial B_{y}}{\partial x}$-$\frac{\partial B_{x}}{\partial y}$$|$\tablenotemark{\emph{2}}\\
  &  current density &   &  \\
J$_{total}$ & Total unsigned vertical & A & J$_{total}$=$\Sigma$$|$$J_{z}$$|$dA\tablenotemark{2}\\
  &  current  &   &  \\
h$_{c}$ & Mean unsigned current & G$^{2}$ m$^{-1}$ & h$_{c}$=$\frac{\mu}{N}$$\Sigma$$|$B$_{z}$J$_{z}$$|$ \\
  &  helicity &   &  \\
h$_{total}$ & Total unsigned current & G$^{2}$ m$^{-1}$ & h$_{total}$=$\Sigma$$|$$h_{c}$$|$ \\
  &  helicity   &   &  \\
$\Psi$ & Mean shear angle & degree & $\Psi$=$\arccos$$\frac{\textbf{B}_{obs}\cdot\textbf{B}_{pot}}{|
 B_{obs}B_{pot}|}$\\
   &   &   &  \\
$\alpha$ & Mean characteristic twist & Mm$^{-1}$ & $\alpha$=$\frac{\mu\Sigma J_{z}B_{z}}{\Sigma B_{z}^{2}}$ \\
   &  parameter  &   &  \\
$\nabla$$B_{h}$ & Mean horizontal gradient & G m$^{-1}$ & $\nabla$$B_{h}$=$\frac{1}{N}$$\Sigma$$\sqrt{(\frac{\partial B_{h}}{\partial x})^{2}+(\frac{\partial B_{h}}{\partial y})^{2}}$\\
  &  of horizontal field &   &  \\
\\\hline
\end{tabular}
\tablenotetext{1}{$\rho_{free}$=$\frac{1}{8\pi}$$|$$\textbf{B}$$_{obs}$-$\textbf{B}$$_{pot}$$|$$^{2}$, where $\textbf{B}$$_{obs}$ and $\textbf{B}$$_{pot}$ are the observed and the potential magnetic fields, respectively. $\textbf{B}$$_{pot}$ was
derived from the observed B$_{z}$ component using the Fourier transform method.}
\tablenotetext{2}{$\mu$ is the magnetic permeability in vacuum (4$\pi$$\times$$10^{-3}$ G m $A^{-1}$).}
\end{table*}

\begin{figure}
	\centering
	\includegraphics[width=17.5cm]{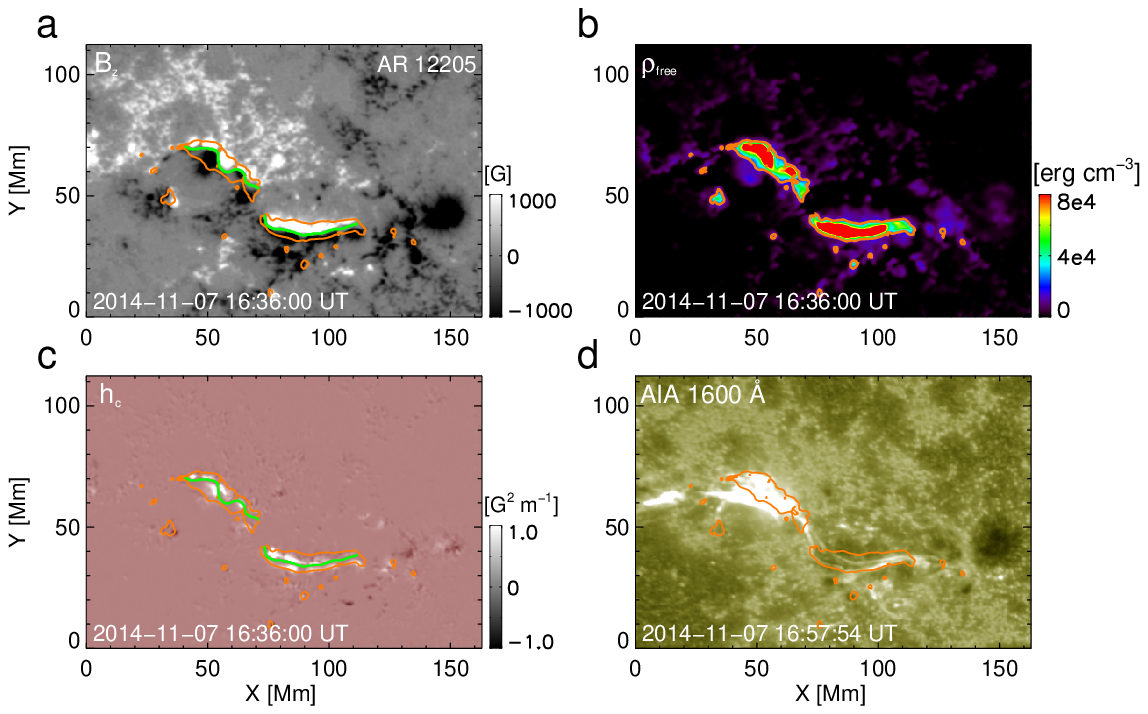}
	\centering
	\caption{Example of an X-class flaring AR showing the distributions of different magnetic parameters. (a)-(c)
Vertical magnetic fields B$_{z}$, photospheric free magnetic
energy density $\rho_{free}$ and current helicity h$_{c}$ of AR 12205, respectively, at 16:36 UT before
the X1.6 flare. (d) The initial flare ribbon of the X1.6 flare observed by SDO at 1600 {\AA} at 16:57:54 UT. Orange contours in (a)-(d) are the areas of $\rho_\mathrm{free}$$\geq$2.0$\times$10$^{4}$ erg cm$^{-3}$.
Green lines in (a) and (c) show the PILs with steep
horizontal magnetic gradient.}
	\label{fig1}
\end{figure}

\begin{figure}
	\centering
	\includegraphics[width=17.5cm]{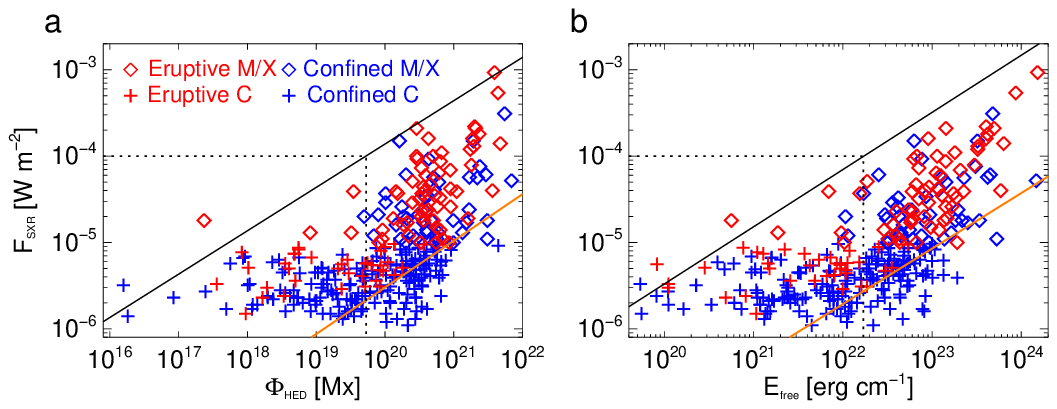}
	\centering
	\caption{Scatter diagrams of flare peak X-ray flux vs. total unsigned magnetic flux $\Phi$$_{HED}$ and total photospheric free energy
density $E_{free}$ within HED region. Red (blue) diamonds are eruptive (confined) M/X-class flares and red (blue) pluses are eruptive (confined) C-class flares. The black straight lines have 99\% of the observed flares below it. Below the orange straight lines, almost all the flares are confined. Black dotted lines show that the X1.0-class flare corresponds to $\Phi$$_{HED}$ of 5.3$\times$$10^{19}$ Mx and $E_{free}$ of 1.7$\times$$10^{22}$ erg cm$^{-1}$.}
	\label{fig1}
\end{figure}

\begin{figure}
	\centering
	\includegraphics[width=17.5cm]{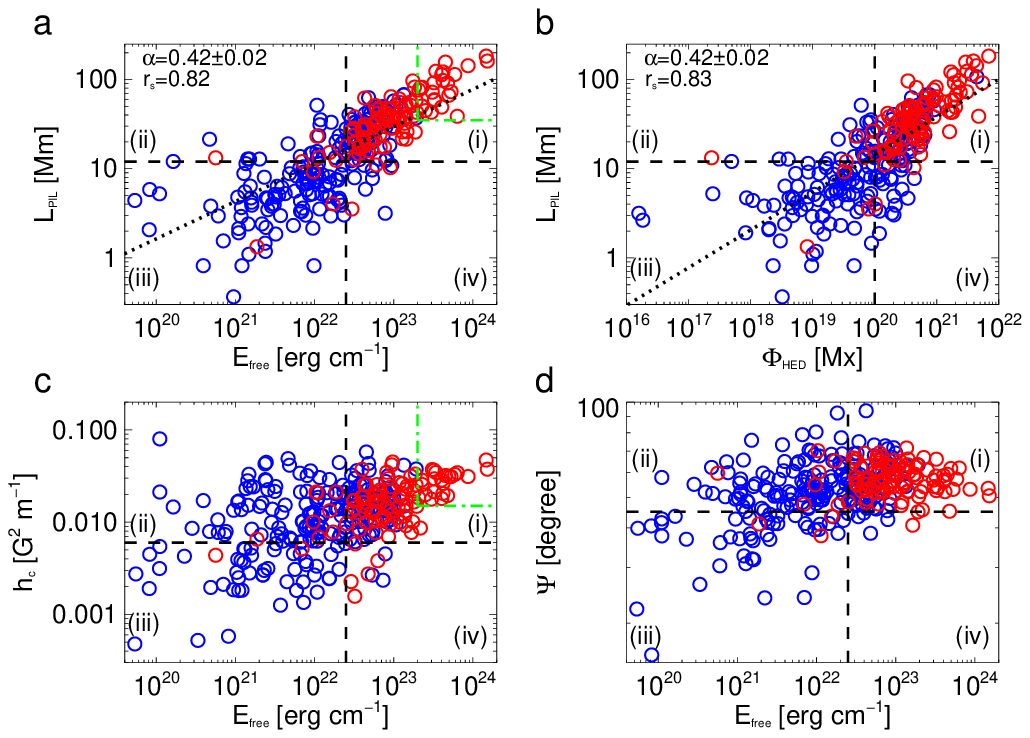}
	\centering
	\caption{Scatter plots of magnetic parameters of C-class and M/X-class flaring ARs. Panels (a)-(d) Diagrams of $E_{free}$$-$L$_{PIL}$, $\Phi$$_{HED}$$-$L$_{PIL}$, $E_{free}$$-$h$_{c}$ and $E_{free}$$-$$\Psi$ for 323 ARs which are calculated about 30 min prior to the flare onset, respectively.
Blue and red circles correspond to 217 C-class and 106 M$/$X-class
flaring ARs, respectively. Black dashed lines indicate the estimated thresholds for E$_{free}$ ($\sim$2.5$\times$$10^{22}$ erg cm$^{-1}$), $L$$_{PIL}$ ($\sim$12 Mm), $\Phi$$_{HED}$ (1.0$\times$$10^{20}$ Mx), h$_{c}$
($\sim$0.006 G$^{2}$ m$^{-1}$) and $\Psi$ ($\sim$45$^{\circ}$) which divide the diagram in four regions, labeled (i), (ii), (iii), and (iv).
Green dash-dotted lines in panels (a) and (c) indicate the estimated thresholds for E$_{free}$ ($\sim$2.0$\times$$10^{23}$ erg cm$^{-1}$), L$_{PIL}$ ($\sim$35 Mm) and h$_{c}$ ($\sim$0.015 G$^{2}$ m$^{-1}$) above which
ARs seem to generate major flares almost exclusively. The dotted
lines in panels (a)-(b) denote the least-squares best logarithmic fits, and slopes
$\alpha$ and Spearman rank order correlation coefficients r$_{s}$
are shown at the top left of each panel.}
	\label{fig3}
\end{figure}

\begin{figure}
	\centering
	\includegraphics[width=17.5cm]{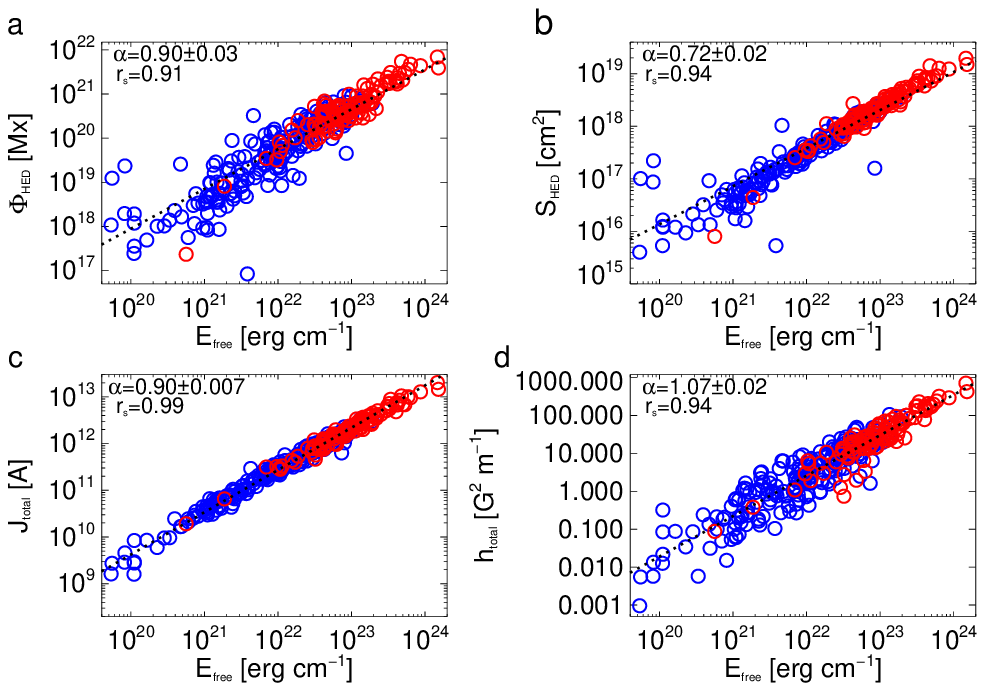}
	\centering
	\caption{Scatter plots of E$_{free}$ vs. $\Phi$$_{HED}$, E$_{free}$ vs. S$_{HED}$, E$_{free}$ vs. J$_{total}$ and E$_{free}$ vs. h$_{total}$.
Blue and red circles correspond to 217 C-class and 106 M$/$X-class
flaring ARs, respectively. Dotted
lines denote the least-squares best logarithmic fits, and slopes
$\alpha$ and Spearman rank order correlation coefficients r$_{s}$
are shown at the top left of each panel.}
	\label{sfig1}
\end{figure}

\clearpage

\end{document}